\documentstyle[prl,aps,twocolumn,graphicx]{revtex}

\begin{document}


\twocolumn[\hsize\textwidth\columnwidth\hsize\csname @twocolumnfalse\endcsname

      \title{Pion abundance and entropy
             in the hydrodynamical description of
             relativistic nuclear collisions}

      \author{Fr\'ed\'erique Grassi and Otavio Socolowski Jr.}

      \address{Instituto de F\'{\i}sica, Universidade de S\~{a}o Paulo, 
               C. P. 66318, 05315-970 S\~{a}o Paulo-SP, Brazil}

      \maketitle

      \begin{abstract}
         We show that a hydrodynamical model with continuous particle
         emission instead of sudden freeze out can explain both the strange
         particle abundances and pion abundance from NA35 without extra
         assumption (e.g., sequential freeze out, modified equation of state,
         sudden plasma hadronization,...). In this scenario, the observation of
         a larger pion abundance is natural and does not imply a higher initial
         entropy and  early plasma phase.
      \end{abstract}


\vskip2pc]


The main purpose of the ongoing and future heavy ion programs at the high
energy laboratories CERN (Switzerland) and Brookhaven National
Laboratory (U.S.A.) is to investigate the formation of hot dense
matter and the possible transition from hadronic matter to quark gluon
plasma. 
Various possible signatures of the appearence of a quark gluon plasma 
(thereafter QGP)
have
been suggested: entropy increase (due to the release of new
degrees of freedom, namely color), 
strangeness increase (due to enhanced strange quark production
 and faster equilibration), 
J/$\psi$ supression (due to color screening),
production of leptons and photons (emitted from a thermalized QGP and
unaffected by strong interactions), etc. These signals
have been studied extensively
experimentally (see for exemple \cite{ha96}).
In this paper, we concentrate on two signatures:
strangeness increase which has been observed between p-p and 
nucleus-nucleus at a fixed energy
and entropy production which is studied via pion measurements.

A
major problem to trace  back any signature unambiguously to a quark gluon phase
is that
it is still unknown which theoretical description describes best  high
energy nuclear collisions.  
On one extreme, one might use a
microscopic model. 
Hadronic microscopic models 
fail
 to reproduce simultaneously 
strange and non-strange particle data in nucleon-nucleon collisions and
central nucleus-nucleus collisions at SPS energies (see \cite{od98,foot1}).
Partonic microscopic models are expected to work at energies higher
 than SPS (however see \cite{ge98} and references therein).

On the other extreme, one might use a thermal or hydrodynamical
model. In such models, it is assumed that a fireball 
(region filled with dense hadronic matter or QGP
 in local thermal and chemical
equilibrium)
is formed 
in a high energy heavy ion collision and evolves.
 Hydrodynamical models have been used successfully to describe 
various kinds of data at
AGS and SPS. In particular they are able to account for
 strangeness data but in their simplest version, fail
to predict big enough pion abundances.

Since as already mentioned,
both strangeness and pion or entropy productions are expected to be
modified by the appearence of a quark gluon plasma, 
looking for 
 a joint explaination (with or without plasma)
 of the relevant data is crucial.
In the case of the thermal and hydrodynamical models mentioned
above, various ways out of the pion problem have
been proposed
(see next section).
In this paper, we study another possibility: use a more accurate emission
mecanism,
namely continuous emission, rather than standard freeze out, in a
hydrodynamical description. 
This explaination has the advantage that no extra 
assumption is needed: once the initial conditions of the
hydrodynamical expansion are fixed, both strangeness and pion yields
come out with the right magnitude.

\noindent
{\bf Hydrodynamical or thermal description with (standard) freeze
  out emission - }
First we remind what is the status of the standard
hydrodynamical or thermal description of relativistic nuclear collisions.
In this kind of description,
hadrons are kept in chemical or thermal
equilibrium until some decoupling criterion has become satisfied.
An exemple of
freeze out criterion often used is that  a certain
 temperature and baryonic potential have been reached.

Since abundances are fixed by the chemical freeze out,
the chemical freeze out parameters can be extracted by analyzing
experimental particle abundances.
This has been done by many groups \cite{so97}. The models have some
variations between them, but as a general rule, while they can
reproduce
strange particle abundances, they
underpredict the pion abundance.
This was first noted by  \cite{da92} in a study of
NA35 data 
and emphasized by
\cite{le93,le95a} in an analysis of the WA85 strange
particle ratios  and EMU05 specific net charge 
$D_q \equiv (N^+-N^-)/(N^++N^-)$ (with $N^+$ and $N^-$, the positive
and negative charge multiplicity respectively).

Various possible mecanisms have been suggested so
that a hadronic gas could yield both the correct strange particle
ratios
and pion multiplicity: sequential freeze out\cite{cl93} or separate
chemical and thermal freeze outs,
hadronic equation of state with excluded volume 
corrections\cite{ri97,ye97,ye98},
non-zero pion chemical potential\cite{da92,ye97},
equilibrated plasma 
undergoing
sudden hadronization and immediate decoupling\cite{le93,le95a,le98}.

All the physical points suggested to salvage the standard
hadronic gas model 
might need to be contemplated
in a precise hydrodynamical description of relativistic nuclear
collisions,
however 
it is somewhat
suprising that one has to go to such kind of details
to reconcile the strange particle and pion data.
{\em Is it not possible to build a simple hydrodynamical model
that yield the correct abundances without extra assumption?}
We discuss this question in the next section\cite{foot3}.

\noindent
{\bf Hydrodynamical description with continuous emission - }
In the standard hydrodynamical models,
one assumes  that the freeze out occurs on a sharp three-dimensional
surface (defined for example by $T(x,y,z,t)=\mathrm{const}$). 
Before crossing it, particles have a hydrodynamical behavior,
and after, they free-stream toward the detectors, keeping memory
of the conditions (flow, temperature) of where and when they crossed the
three dimensional surface.
The Cooper-Frye formula 
\cite{co74} gives the invariant momentum distribution
in this case
\begin{equation}
E d^3N/dp^3=\int_{\sigma} d\sigma_{\mu} p^{\mu} f(x,p). \label{CF} 
\end{equation}
$d\sigma_\mu$ is a normal vector to the freeze out surface $\sigma$
and $f$ the distribution function of the type of particles considered. 
This is the formula implicitly used in all standard thermal and 
hydrodynamical model
calculations of the previous section.

The notion that particle emission
 does not necessarily occur on a
three dimensional surface but may be continuous was incorporated in
a hydrodynamical description in
 \cite{gr}.
In this model,
the fluid is assumed to have two components, 
a free part plus an interacting part and 
its distribution function reads
\begin{equation}
f(x,p)=f_{\mathrm free}(x,p)+f_{\mathrm int}(x,p).
\end{equation}
$f_{\mathrm free}$ counts all the particles that 
last scattered earlier at some point and
are at time $x^0$ in $\vec{x}$.
$f_{\mathrm int}$ describes all the particles that are still 
 interacting (i.e., that will
suffer collisions at time $>x^0$ and change momentum).
The invariant momentum distribution is then
\begin{equation}
E d^3N/dp^3=\int d^4x\,
D_{\mu} [p^{\mu} f_{\mathrm free}(x,p)]. \label{CE}
\end{equation}
$D_{\mu} [p^{\mu} f_{\mathrm free}(x,p)]$ is a covariant 
divergence in general coordinates and 
$d^4x$ is the invariant volume element.
 A priori formula (\ref{CE}) is sensitive to the whole fluid history
 and not just to freeze out conditions as in formula (\ref{CF}).

To compare particle abundances in the continuous emission and freeze out scenarios,
 we  use a simplified
framework to describe the fluid expansion, 
namely we suppose longitudinal expansion
only and longitudinal boost invariance\cite{bj83}.
This approximation allows to carry out
 some 
 calculations analytically and turns the 
physics involved more transparent.
It is implicit however that this description applies at best to the
midrapidity
region.
We will therefore concentrate on midrapidity abundances, precisely data
 from NA35.

In this simplified framework,
in the case of a fluid with 
freeze out at a constant temperature and chemical potential, 
the Cooper-Frye formula (\ref{CF}) can
be rewritten ignoring transverse expansion as \cite{ru87}
\begin{eqnarray}
\left. \frac{dN}{dy p_{\perp} dp_{\perp}}\right|_{y=0} &&=\frac{gR^2}{2 \pi} 
\tau_{\mathrm fo}\left(T_{\mathrm fo},T_0,\tau_0\right) m_{\perp} \nonumber \\
 \times && \sum_{n=1}^{\infty} (\mp)^{n+1} \exp\left(\frac{n \mu_{\mathrm fo}}{T_{\mathrm fo}}\right)
K_1\left(\frac{n m_{\perp}}{T_{\mathrm fo}}\right). \label{CF_Bj}               
\end{eqnarray}
(The plus sign corresponds to bosons and minus, to fermions.)
It depends  on the conditions at freeze out: $T_{\mathrm fo}$ and 
$\mu_{\mathrm fo}=\mu_{b\,\mathrm{fo}}B+\mu_{S\,\mathrm{fo}}S$, 
with $B$ and $S$ the baryon number
and strangeness of the hadron species considered, and 
$\mu_{S\,\mathrm{fo}}(\mu_{b\,\mathrm{fo}},T_{\mathrm fo})$ 
obtained by imposing strangeness
neutrality. 
So the experimental spectra of particles teach us in that case
 what the conditions were at freeze out.

For continuous emission,
we can approximate equation (\ref{CE}) as
\cite{gr}
\begin{eqnarray}
&& \left. \frac{dN}{dy p_{\perp} dp_{\perp}}\right|_{y=0} \approx \frac{2 g}{(2 \pi)^2} \nonumber \\
&& \times \int_{{\cal P}=0.5} d\phi  d \eta   
\frac{m_{\perp} \cosh \eta \tau_F \rho d\rho       
+ p_{\perp} \cos \phi \rho_F \tau d\tau }
{\exp\left[(m_{\perp}\cosh \eta - \mu) /T\right] \pm 1}, 
\label{CE_Bj}
\end{eqnarray}
where
${\cal P}$  is the probabality to escape without collision calculated
with a Glauber formula,
$\tau_F$
(resp. $\rho_F$)
is solution of 
${\cal P}(\tau_F,\rho,\phi,\eta;v_{\perp})=0.5$
(resp. ${\cal P}(\tau,\rho_F,\phi,\eta;v_{\perp})=0.5$).
In (\ref{CE_Bj}), various $T$ and $\mu=\mu_b B + \mu_S S$ appear
(again $\mu_S$ is obtained from strangeness neutrality), 
reflecting the whole fluid history,
not just $T_{\mathrm fo}$ and $\mu_{b\,\mathrm{fo}}$.
This history is known by solving the hydrodynamical equations of a
hadronic gas with continuous emisison. It depends only on the initial
conditions $T_0$ and $\mu_{b0}$. Therefore (\ref{CE_Bj}) only depends
on the initial conditions.
\begin{figure}
\centering\includegraphics*[scale=0.3]{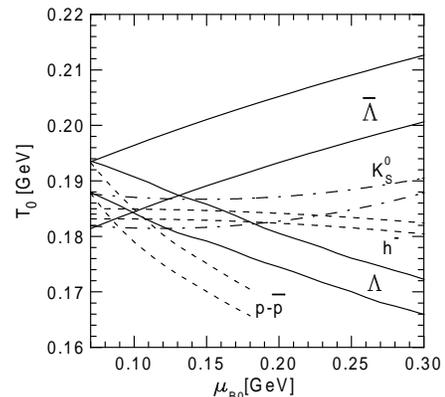}
\caption{Allowed region for the initial conditions determined from the NA35
data.}
\end{figure}

Once the spectra are known, they can be integrated to get abundances.
Figure 1 shows the allowed region of initial conditions that lead to
the experimental NA35 midrapidity values \cite{NA35}
$\Lambda=1.26 \pm 0.22_{p_\perp > 0.5 \mathrm{GeV}}$,
$\overline{\Lambda}= 0.44 \pm 0.16_{p_\perp > 0.5 \mathrm{GeV}}$,
$K_S^0= 1.30 \pm 0.22_{p_\perp > 0.62 \mathrm{GeV}}$,
$h^-=27 \pm 1$ and $p-\overline{p}= 3.2 \pm 0.4 $.
We do not use the $K^+$ and $K^-$ abundances because they were  measured 
outside the mid-rapidity region.
For heavy particles, initial conditions 
dominate\cite{grs}
so when looking at their abundances in  finite $p_{\perp}$ windows,
transverse flow (assumed zero initially) can be neglected as we did.
For negatives, the experimental abundance is for all $p_{\perp}$
so tranverse flow does not affect their abundance either.
The allowed
window corresponds to $T_0 \approx 185$ MeV, $\mu_{b0}\approx 100$ MeV
, for an ideal gas equation of state and 
the strangeness saturation factor $\gamma_s=1.3$

Using a more sofisticated equation of state, the value of $T_0$
might be decreased \cite{grs} by some 10-15\%, i.e., to 155-165 MeV,
compatible with  (i.e., below) QCD lattice values for the phase transition
temperature from QGP
to hadronic matter.
Our value of $\gamma_S$ is above 1 and this might looked surprising.
However, its value is decreased by some 15\% when looking at a more realistic
equation of state. In addition, we have imposed  strangeness
neutrality, it is possible that this is a too strong constraint when
analyzing data taken in a very restricted rapidity region (see
\cite{so94} where a similar problem was encountered).Using a larger value
of $\gamma_s$, the size of the allowed window for initial 
conditions in figure 1 increases.
There are other factors
that influence the precise
location and size of the window: values of the cross section 
needed to compute $\cal P$ (taken constant and equal for simplicity
here), value of the cutoff in ${\cal P}=0.5$, etc.
However the important point is that it is
possible to find initial conditions of the hydrodynamical expansion
such that strange and non-strange particle abundances can be
reproduced
simultaneously without extra assumption.

To illustrate why the continuous emission is able to reproduce both
strangeness and pion data,
in table 1, we compare results from the continuous emission scenario
and the freeze out model. We took $T_0=185$ MeV and
$\mu_{b0}=100$ MeV for the continuous emission case, $T_{\mathrm fo}=185$ MeV and
$\mu_{b\,\mathrm{fo}}=100$ MeV for the freeze out case. 
We used $\gamma_s=1.3$ for both\cite{foot2}.
We expect roughly
similar results for both models for heavy particles:
in the continuous
emission case, due to thermal suppression, they are mostly emitted
early\cite{grs}, i.e., in similar conditions than in
 the freeze out model. Pions in
the freeze out case are too few as  discussed previously.
 Pions in
the continuous emission case, on
the other side, are emitted early and then on, so we expect to have
more of them. This is precisely what we see in the 
table.

The initial conditions that we discussed so far correspond to a hadron
gas, starting its hydrodynamical evolution. However the present 
continuous emission scenario is in fact probably compatible
with the possibility that a QGP was created before for the following reason. 
Continuous emission is
possible from a QGP but is inhibited by two factors: 1) only color
singlet objects (not single quarks) can escape from it 2) the QGP core
is surrounded by a dense hadronic region that the color singlets would have
to cross {\em without} collisions to modify the previous
results. 
In this case, particle emission would occur mostly in the hadronic phase.
Numerical estimates are however necessary to back up these qualitative
arguments.
\begin{figure}
\centering\includegraphics*[scale=0.8]{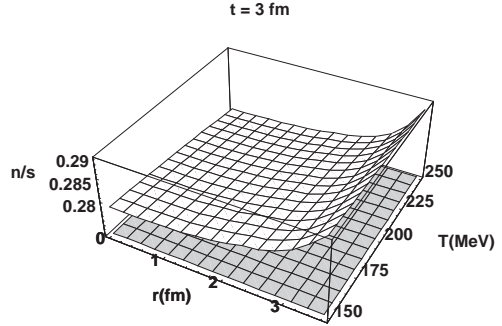}
\caption{Massless pion gas:
value of the pion density over entropy density as function of the
initial temperature and position for $t=3\mathrm{\;fm}$ in the continuous emission scenario 
(white surface). For comparison the value for the
freeze out case is shown (gray surface).}
\end{figure}

\noindent
{\bf Relation between pion number and entropy - }
In a hydrodynamical model without shocks and dissipation, entropy is 
conserved. In the usual freeze out scenario, this is an important
point because the initial entropy can be determined from the final 
multiplicity.
To illustrate this connection, let us consider a massless pion
gas. Statistical mechanics yields the following relationship for the
entropy density and the pion density $s=3.6 n_{\pi}$.
In the longitudinal boost invariant model, the entropy conservation
equation gives: $s \tau = \mathrm{const}$. Using $dV=\tau dy \pi R^2$ at $y=0$,
this can be written: $dS(\tau)/(\pi R^2 dy)=\mathrm{const}$. 
So we can rewrite
\begin{equation}
\frac{dS}{dy}(\tau)=\frac{dS}{dy}=3.6 \frac{dN_{\pi}}{dy}.
\end{equation}
Therefore, experimental knowledge of  $dN_{\pi}/dy$, 
permits to extract $dS/dy$, which are both independent of time.

In the continuous emission case, $ s \neq 3.6 n_{\pi} $
because the distribution function needed to compute the pion density
now 
depends on the escape probability
$\cal{P}$, embodied in $f_{\mathrm free}$ (cf. eq. (2)).
In fact, using similar methods than in \cite{gr}, one can show that
\begin{equation}
s=\left[1+\frac{3 \beta - \alpha}{4(1+\alpha)}\right]  3.6 n_{\pi},
\end{equation}
where  
$
\alpha(t,\rho)=\left(4\pi\right)^{-1} \int d\phi  d\theta \sin \theta 
\left[{\cal P}/(1-{\cal P})\right]_{z=0}$ 
and
$\beta(t,\rho)=\left(4\pi\right)^{-1} \int d\phi  d\theta \sin \theta \cos^2\theta
\left[{\cal P}/(1-{\cal P})\right]_{z=0}$ .
This relationship  is plotted in figure 2. For all radii $\rho$, time
$\tau$ and inicial condition $T_0$, $s(\tau,\rho) \leq 3.6 n_{\pi}(\tau,\rho)$.
One then gets in the longitudinal boost invariant model:\\
\begin{equation}
\frac{dS}{dy} \leq 3.6 \frac{dN_{\pi}(\tau)}{dy}, 
\end{equation}
in other words,  for a fixed entropy, there
are more pions in the continuous emission case than in the freeze out
case
(for all times $\tau$).
This is in fact
expected (even using a more realistic equation of
state and including pions from decays):
since pions are emitted continuously,
there are more copious than in the the usual freeze out case for a
given $dS/dy$ (itself fixed by the initial conditions).

As a consequence, a large experimental value of $dN_{\pi}/dy$ should
not  necessarily be associated to a large $dS/dy$ and be considered
a hint of  QGP formation as usually done in freeze out models\cite{le93}.
A larger
 $dN_{\pi}/dy$ and not large  $dS/dy$
is a natural consequence of continuous emission, compared to freeze out.
This result is in contrast for exemple with \cite{le93,le95a}.

\noindent
{\bf Conclusion - }
In this paper, we discussed data on strange and non-strange particles
by NA35, from an hydrodynamical point of view. 
The standard  model with sudden freeze out can reproduce the strange
particle data but underpredicts the pion abundance, if no extra
assumption is made.
We showed that a hydrodynamical model with a more precise emission
process, continuous emission, can reproduce both the strange and
non-strange particle data without extra assumption.

This indicates the necessity of doing a more accurate description of
particle emission in hydrodynamics, else some problems might
artificially appear, such as a too low predicted  pion abundance
as discussed here or a too high freeze out density \cite{grs}.

This point is reinforced by the fact that a large pion number is
usually associated with a large entropy and  QGP formation. Here we
showed that a large pion number can be generated by continuous emission
without modifying the
entropy. (The larger pion emission will cause a faster cooling and
shorter fluid lifetime.) In other words, a better understanding of
particle emission in the hydrodynamical regime
is also necessary to assess the possibility of QGP formation in
relativistic heavy ion collisions.

There is a growing tendency to use hydrodynamical  models to describe
relativistic nuclear
collisions, there is also a growing concern to modelize
freeze out better \cite{du99,ba99}. However it seems that the very notion of
particle emission during the hydrodynamical expansion needs to be put
under more scrutiny.

This work was partially supported by FAPESP (proc. 98/14990-0,
98/2249-4
and 99/0529-2)
and CNPq (proc. 300054/92-0).

\begin{table}
\caption{Comparison of experimental particle abundances with
continuous emission and freeze out predictions.}
\begin{center}
\begin{tabular}{cccc}
& Experimental value & Continuous emission & Freeze out \\ \hline
$\Lambda $ & 1.26$\pm $0.22 & 1.05 & 0.98 \\ 
$\bar \Lambda $ & 0.44$\pm $0.16 & 0.28 & 0.46 \\ 
$p-\bar p$ & 3.2$\pm $0.4 & 3.6 & 1.52 \\ 
$h^{-}$ & 27$\pm $1 & 27.6 & 16 \\ 
$K_S^0$ & 1.30$\pm $0.22 & 1.27 & 1.08 \\ 
\end{tabular}
\end{center}
\end{table}


\begin{references}
\bibitem{foot2} In addition, there is an overall
multiplicative factor in (4), $\tau_{\mathrm fo}$,
 which we set to 1. If increased to 1.3, all the
 abundances except the pion one and $p-\bar{p}$ 
are still  described. 
If increased more, $\bar{\Lambda}$ is not
reproduced anymore.
\end{references}
\end{document}